\documentclass[11pt]{revtex4-1}
\usepackage{graphicx}
\usepackage{dcolumn}
\usepackage{color}
\usepackage{bm}
\usepackage{amsmath}
\usepackage{amsfonts}

\usepackage{filecontents}

\input epsf

\begin{document}

\title{Interacting Dark Energy in $ f(T) $ cosmology : A Dynamical System analysis}

\author{Sujay Kr. Biswas\footnote{sujaymathju@gmail.com}}
\affiliation{Department of Mathematics, Ramakrishna Mission Vivekananda Centenary College, Rahara, Kolkata-700 118, West Bengal, India.}

\author{Subenoy Chakraborty\footnote{schakraborty@math.jdvu.ac.in}}

\affiliation{Department of Mathematics, Jadavpur University, Kolkata-700 032, India.}{}


\begin{abstract}
  The present work deals with an interacting dark energy model in the framework of $ f(T) $ cosmology. A cosmologically viable form of $ f(T) $ is chosen ( $ T $ is the torsion scalar in teleparallelism ) in the background of flat homogeneous and isotropic Friedmann-Robertson-Walker ( FRW ) space-time model of the universe. The matter content of the universe is chosen as dust interacting with minimally coupled scalar field. The evolution equations are reduced to an autonomous system of ordinary differential equations by suitable transformation of variables. The nature of critical points are analyzed by evaluating the eigenvalues of the linearized Jacobi matrix and stable attractors are examined from the point of view of cosmology. Finally, both classical and quantum stability of the model have been discussed.\\



\end{abstract}

\maketitle
\section{Introduction}

 The prediction through the SNeIa Hubble diagram [1,2] that the universe is at present going through on accelerating phase, has been confirmed by wide range of data, from more recent SNeIa data to BAOs and CMBR anisotropies [3-7]. However, the overwhelming abundance of observational evidences for cosmic speed up does not match with standard cosmology in the frame work of General Relativity (GR). To resolve this paradox, an unexpected new ingredient in the form of a negative pressured component has been introduced but it poses difficult questions on its nature and introduces further problems hard to be solved. So it is naturally speculated that these observational evidences may be the first signal of a breakdown of our understanding of the laws of gravity on cosmological scale. As a result several modified gravity theories has been developed of which $f(R)$-gravity theory[8,9] gets much attention. In this theory the scalar curvature $ R $ is replaced by a suitably chosen function $ f(R) $ in the gravity Lagrangian.\\

An equivalent formulation is represented by teleparallelism where instead of curvature, torsion is responsible for the gravitational interaction [10-12]. This model was first proposed by Einstein  for unifying electromagnetism and gravity on Weitzenb\"{o}ck non-Riemannian manifold. The Weitzenb\"{o}ck connection replaces the Levi-Civita one on the underlying Riemann-Cartan space-time. Thus gravitational interaction is not purely geometrical rather, the torsion acts as a force, allowing the interpretation of gravity as a gauge theory of the translation group[13]. Although, conceptually, GR and teleparallel gravity are distinct but they yield equivalent dynamics at least at the classical level.\\

In analogy with $f(R)$-gravity, a generalization to teleparallel gravity is obtained by replacing $ T $ ( the torsion scalar ) with a generic function $ f(T)$. A particular important consequence is the breakdown of the equivalence with the classical GR and hence predicts different dynamics [14-16]. Further, modified teleparallel gravity preserves the field equations to be still second order in the field derivatives in contrast to fourth order equations in $ f(R) $-gravity. However, this modified theory suffers from the lack of Local Lorentz Invariance (LLI), hence all the sixteen components of the vierbien are independent and one can not fix six of them by a gauge choice[17].

 For the past few years, models based on DE interacting with dark matter ( DM ) or any other exotic matter components have gained great impetus. Such interacting DE models can successfully explain numerous cosmological puzzles namely, phantom crossing, cosmic coincidence and cosmic age problem [18-23].

 The present paper is devoted to the study of dynamics of interacting dark energy in $ f(T) $ cosmology. The interaction between DE and DM could be a major issue to be confronted in studying the physics of DE. However, due to the nature of these two components remaining unknown, it will not be possible to derive the precise form of the interaction from first principles. One has to assume a specific coupling from the outset or determine it from phenomenological requirements. Further, in the framework of field theory it is natural to consider the inevitable interaction between the dark components. An appropriate interaction between DE and DM can provide a mechanism to alleviate the coincidence problem. Moreover, complementary observational signatures of the interaction between DE and DM have been obtained from the cosmic expansion history by using WMAP, SNIa, BAO and SDSS data as well as the growth of cosmic structure. Interestingly, it was disclosed that the late integrated Sachs-Wolfe effect has the unique ability to provide insight into the coupling between the dark sectors. Further, in view of the continuity equations, the interaction between DE and DM  must be a function of the energy densities multiplied by a quantity with units of the inverse of time which has  the natural choice as Hubble parameter.

 Usually one assumes a phenomenological form of interaction between matter and dark energy which are the dominant components of the cosmic composition  [24-28] today. In fact, the  interaction term gives the rate of exchange of energy density in the dark sector.

 The motivation for choosing such a complex system is to test of such a weird model could explain the overall cosmological evolution. Due to complicated field equations, we study the dynamics in the phase space associated to this scenario around both hyperbolic and non-hyperbolic critical points. Considering first order perturbations near the critical points, we examine the nature of critical points by considering the eigenvalues of first order perturbed matrix. Previously, similar dynamical system analysis has been done by us in Dirac-Born-infeld gravity theory [29] and in Brane Scenario [30]. In DBI model, critical points are evaluated, analyzed and their stability have been discussed. In brane gravity model which is essentially on higher dimensional theory, in addition to the above analysis the classical stability of the model itself has been examined. In the present $ f(T)$ gravity model, choosing the function '$ f $' appropriately the evolution equations are converted to an autonomous system by suitable choice of the auxiliary variables. In addition to the above course of analysis in the present work we shall also examine the stability of the system both classically as well as quantum mechanically.

 We follow the plan : In section II, we present the basic theory of $ f(T) $ gravity. In section III, we introduce the basic equations and formation of dynamical system. While the critical points of the dynamical system and phase space analysis to the corresponding critical points are shown in section IV. In section V and VI, we work out the stability analysis of the  critical points. The cosmological implications at the critical points are discussed in section VII. We conclude and summarize in section VIII. Throughout the paper we use natural units ( $ 8\pi G=\frac{8 \pi}{m_{PL}^{2}}=\hbar=c=1 $ ). \\
\section { $f(T)$- gravity theory }

 A possible way to modify gravity beyond general relativity is to consider the Weitzenb\"{o}ck connection, which has no curvature but torsion, rather than the curvature defined by the Levi-Civita connection. Such an approach is termed as 'teleparallelism'[12,31-33], which was also considered by Einstein himself[10,11]. However, in the context of the present accelerated expansion of the universe, the teleparallel Lagrangian density described by the torsion scalar $ T $ has been extended to a function of $ T $  ( $ f(T) $ ) [34,35], equivalent to the concept of $ f(R) $ gravity.

 In this new modified gravity theory ( called $ f(T) $-gravity ), the gravity is no longer caused by curved space-time but torsion and moreover, the field equations are only second order unlike the fourth order equations in the $ f(R) $ theory.

 In $ f(T) $ gravity theory the action is written as
 \begin{equation}
  I=\frac{1}{2k^{2}} \int d^{4}x [|e|(T + f(T)) + L_{m}]
 \end{equation}

 Where $ T $ is the torsion scalar, $ f(T)$ is a differentiable function of the torsion, $ L_{m} $ corresponds to the matter Lagrangian, $|e|=det(e^{A}_{\mu})=\sqrt{-g}$ and  $ k^{2}=8\pi G $. The torsion scalar is defined as
  \begin{equation}
  T=S_{\rho} \ ^{\mu\nu}T^{\rho} \ _{\mu\nu}
 \end{equation}
 with
  \begin{equation}
  S_{\rho} \ ^{\mu\nu}= \frac{1}{2}(K^{\mu\nu} \ _{\rho} + \delta^{\mu} _{\rho} T^{\theta\nu} \ _{\theta}-\delta^{\nu} _{\rho} T^{\theta\mu} \ _{\theta} ),
 \end{equation}

 \begin{equation}
  K^{\mu\nu} \ _{\rho}= -\frac{1}{2}(T^{\mu\nu} \ _{\rho}-T^{\nu\mu} \ _{\rho}-T_{\rho} \ ^{\mu\nu})
  \end{equation}

  \begin{equation}
  T^{\lambda} \ _{\mu\nu}= \stackrel{\bf w} {\Gamma}^{\lambda}\ _{\nu\mu} - \stackrel{\bf w} {\Gamma}^{\lambda}\ _{\mu\nu} = e^{\lambda}_{A}(\partial_{\mu}e^{A}_{\nu}- \partial_{\nu}e^{A}_{\mu})
 \end{equation}
 In teleparallelism, orthogonal tetrad components  $ e_{A}(x^{\mu}) $ are used as dynamical objects and are cosidered as an orthonormal basis for the tangent space at each point $ x^{\mu} $ of the manifold. So we have
  $$ e_{A}e_{B}=\eta_{AB}= diag(1, -1, -1, -1) $$.
  Here each vector $ e_{A} $ can be described by its components $e^{\mu}_{A}$ , $ \mu$ =0,1,2,3 ($A$ =0,1,2,3) in a coordinate basis i.e. $ e_{A} = e^{\mu}_{A}\partial_{\mu} $. Here capital letters refer to the tangent space while greek indices label coordinates on the manifold. The metric tensor is obtained from the dual vierbein as
   \begin{equation}
    g_{\mu\nu}(x) = \eta_{AB} e^{A}_{\mu}(x)e^{B}_{\nu}(x)
  \end{equation}
  In the above $T^{\lambda} \ _{\mu\nu}$ is the curvature less Weitzenb\"{o}ck connection and it encompasses all the information about the gravitational field. The contorsion tensor  $K^{\mu\nu} \ _{\rho}$ gives the difference between Weitzenb\"{o}ck and Levi-Civita connections.

  It is straight forward to show that this equation of motion reduces to Einstein gravity when $ f(T)= 0 $. Indeed, it is the equivalency between the teleparallel theory and Einstein gravity [12]. The theory has been found to address the issue of cosmic acceleration in the early and late evolution of universe [36] but this crucially depends on the choice of $ f(T) $.  For instance, exponential or power-law form of $ f(T) $ can not lead to phantom crossing [37]. But subsequently, it has been shown that crossing of the phantom divide line for the effective equation of state of two specific models in $ f(T) $ gravity is possible[38] and the best fit results suggest that the observations favour a crossing of the phantom barrier. Then reconstruction of $ f(T) $ models has been reported in [39,40] while a detailed cosmological analysis is performed in [41,42] and thermodynamics of $ f(T) $ cosmology [43] including the generalized 2nd law of thermodynamics has been investigated.\\

  Further, it should be noted that a constant $ f(T) $ acts like as a Cosmological Constant while $ f $ linear in $ T $ (i.e $ f_{T}=constant $ ) is simply a redefinition of Newton's Constant $ G $. A desirable choice of $ f(T) $ is such that at high redshift general relativity should hold so as to agree with primordial nucleosynthesis and cosmic microwave background constraints i.e $ f/T \rightarrow 0 $  at early times ($ a\ll 1 $ ) [44]. On the other hand, in the asymptotic future there will be de Sitter state. A simplest choice of $ f $ is the polynomial form [44]

 \begin{equation}
  f(T)=\beta (-T)^{n}
  \end{equation}
  where $\beta $ is arbitrary constant.

  In particular for $ n=\frac{1}{2} $ the model gives the same expansion history as DGP gravity [44,45] and hence $ f(T) $ gravity can be considered to be related to higher dimension theories. Note that the restriction $ 'n\ll 1' $ gives $ f(T) $ a viable model compared to current observed data set and also puts the rescale factor $ f_{T} $ to Newton's constant to be small. The effective DE equation of state varies from $ \omega = -1+n $ in the past to $ \omega=-1 $ in the future.

  Throughout the work we consider the flat homogeneous and isotropic FRW universe with the metric
  \begin{equation}
   ds^{2} =dt^{2} - a(t)^{2}\sum(dx^{i})^{2},
  \end{equation}
  where t is cosmic time. For this metric

  $$ e^{A}_{\mu}=diag (1,a(t),a(t),a(t)) ,$$
  where  $ a(t) $ is cosmological scale factor. By combining with (5), (3) and (4) one obtains

  $$ T=S^{\rho\mu\nu} T _{\rho\mu\nu}= -6H^{2} , $$
  where $  H=\frac{\dot{a}}{a}  $ is the Hubble parameter and dot stands for derivative with respect to t.
   \\

  \section{Basic Equations in $ f(T) $ cosmology and formation of Dynamical system :}

  In $f(T)$ gravity theory, we consider flat, homogeneous and isotropic FRW space-time as the model of our universe and the matter is chosen as dark matter in the form of dust(having energy density $\rho_{m}$) interacting with the dark energy which is chosen as a minimally coupled scalar field $\phi$ having self interacting potential $V(\phi)$. So the energy density and pressure corresponding to this scalar field are
  \begin{equation}
    \rho_{\phi}=\frac{1}{2}\dot{\phi}^{2} + V(\phi) , p_{\phi} = \frac{1}{2}\dot{\phi}^{2} - V(\phi)
 \end{equation}
 Thus the modified Friedmann equations in $f(T)$- gravity are[14,15]

\begin{equation}
 H^{2} = { \frac{1}{ (2f_{T}+1) }}[  \frac{1}{3}(\rho_{\phi}+\rho_{m}) -\frac{f}{6} ]
\end{equation}

\begin{equation}
   \dot{H} = - \frac{1}{2} [ \frac{\rho_{m} + \rho_{\phi} + p_{\phi}}{1+ f_{T}+ 2Tf_{TT}} ]
\end{equation}
The energy balance equations for the individual dark components are
\begin{equation}
    \dot{\rho_{m}} +3H\rho_{m}=Q
\end{equation}
and

\begin{equation}
    \dot{\rho_{\phi}} +3H(1+\omega_{\phi})\rho_{\phi} =-Q
\end{equation}
 Where the interaction term  $ Q $ corresponds to energy exchange between dark energy and dark matter. The positivity of $Q$ indicates a transfer of energy from dark energy to dark matter. This is required to alleviate the coincidence problem and is compatible with the 2nd. law of thermodynamics. For the time being $Q$ is unspecified, only it is assumed that $Q$ does not change sign during the cosmic evolution. Due to the unknown nature of the two dark components (DM and DE), the precise form of the interaction can not be determined from the outset or from phenomenological requirements. However, it is speculated that the interaction term may lead to a major issue to be confronted in studying the physics of DE. Also, from the view point of field theory, interaction between the dark components appears naturally. Moreover, an appropriate interaction between the DE and DM can provide a mechanism to alleviate the coincidence problem. Further, in view of the continuity equations ( i.e. equations (12) and (13) ) the interaction between DE and DM should be a function of energy densities multiplied by a factor having dimension inverse of time and Hubble parameter is a natural choice for it. Thus phenomenologically, $Q$ can be chosen as (i) $ Q=Q( H\rho_{m} ) $ ,  (ii) $ Q=Q( H\rho_{\phi} ) $, (iii) $ Q=Q[H(\rho_{\phi}+\rho_{m})] $   or more generally (iv) $ Q=Q( H\rho_{\phi},H\rho_{m} )$.
 In the present work, for simplicity $Q$ is chosen as $Q=\alpha H \rho_{m}$, where the coupling parameter $'\alpha'$ is assumed to be small.

 Now, using (9) in the continuity equation (13) the evolution of the scalar field is given by

 \begin{equation}
    \ddot{\phi} + 3H\dot{\phi} + \frac{dV(\phi)}{d\phi}= -\frac{Q}{\dot{\phi}}
 \end{equation} (the modified Klein-Gordon equations)\\

 Due to complicated form of the evolution equations, it is not possible to have analytic solution so to have a qualitative idea about the cosmological behavior we shall put the evolution equations into an autonomous dynamical system. For this, we introduce the new variables

 \begin{equation}
    x=\frac{\dot{\phi}}{\sqrt{6}H}  ,  y= \frac{\sqrt{V(\phi)}}{\sqrt{3}H}  ,  \Omega_{m}= \frac{\rho_{m}}{3 H^{2}}
 \end{equation}

 which are normalized over Hubble scale. $ \Omega_{m} $ is the density parameter for DM. As a result, the evolution equations reduce to the following autonomous system of ordinary differential equations ( after some algebra )\\\\

  $$\frac{dx}{dN}=(1+ x^{2}-y^{2})(\frac{3}{2}x-\frac{\alpha}{2x}) + (\alpha-3)x-\sqrt{\frac{3}{2}}y^{2}\lambda $$
   $$\frac{dy}{dN}=y[ \sqrt{\frac{3}{2}}x\lambda + \frac{3}{2}(1+x^{2}-y^{2})] $$
   \begin{equation}
   \frac{d\Omega_{m}}{dN} = \Omega_{m}[\alpha + 3(x^{2}-y^{2}) ]\\
   \end{equation}

 where we choose $ \frac{V'(\phi)}{V(\phi)} =\lambda $, a constant such that $ V(\phi)= e^{\lambda\phi} $ and the independent variable is chosen as             $ N=\ln a $, which is called the e-folding parameter. The system of equations in (16) is analyzed by first equating them to zero to obtain the critical points. Next we perturb equations up to first order about the critical points and check their stability.\\

 In deriving the above autonomous system of ODE we choose $ f(T)=\beta \sqrt{-T} $ (i.e $ n=\frac{1}{2}$ in equation (7) ) and the interaction is chosen in the form $ Q= \alpha H \rho_{m} $. As mentioned earlier, the positive coupling is very reassuring in view of coincidence problem. However, models with negative coupling parameter (indicates decay of DM into DE ) allow for the possibility that there is no DE field in the very early universe and that DE 'Condenses' as result of the slow decay of DM [46]. Also, it has been shown [47] that the coupling parameter is weakly constrained to negative values by Planck measurements. Further, the negative coupling can not be counted to resolve the tension between the Planck and HST measurements of the Hubble parameter [47]. Although, the negative coupling does not help to alleviate the coincidence problem, it appears in the observed data fittings that models with negative coupling shows most significant departure from zero coupling.

 Further, using the normalized variables in the first modified Friedmann equation, we obtain the density parameter for the dark matter as
 \begin{equation}
    \Omega_{m}=1-x^{2} -y^{2}
\end{equation}
 Due to the energy condition $ 0< \Omega_{m} <1 $ , so for fix $ \Omega_{m} $ , $ (x,y) $ lies on the circle  $ x^{2} + y^{2} =1-\Omega_{m} $ .
 In fact, the phase space $ ( x,y,\Omega_{m} ) $ of the autonomous system  (16) forms a paraboloid ( $ x^{2}+y^{2} +\Omega_{m}=1 $ ) bounded by            $ \Omega_{m} =0 $ and $ \Omega_{m} =1 $. Hence the phase space is finite .

 The cosmological parameters related to the scalar field namely the equation of state parameter $ \omega_{\phi} $ and the density parameter                $ \Omega_{\phi} $ can be expressed by newly defined variables as
 \begin{equation}
    \omega_{\phi} = \frac{p_{\phi}}{\rho_{\phi}} =\frac{x^{2}-y^{2}}{x^{2}+y^{2}} ,     \Omega_{\phi} = \frac{\rho_{\phi}}{3H^{2}} = x^{2} + y^{2}
\end{equation}
 and
 \begin{equation}
     \omega_{Tot} =\frac{p}{\rho} = \frac{p_{\phi}}{\rho_{\phi}+\rho_{m}}= \frac{x^{2}-y^{2}}{x^{2}+y^{2}+\Omega_{m}}= x^{2}-y^{2}
 \end{equation}
 since $ x^{2}+y^{2} +\Omega_{m}=1 $.
 Also, the deceleration parameter has the explicit form :
 \begin{equation}
    q= -1-\frac{\dot{H}}{H^{2}}= -1 + \frac{3}{2}(2x^{2} + \Omega_{m})
 \end{equation}

 and the equation of state parameter of the equivalent geometric matter is

\begin{equation}
    \omega_{g}=\frac{(q-2)}{3}
 \end{equation}
 \section{Critical points and Phase -Space Analysis :}

 We analyze the stability of the corresponding dynamical system about the critical points. We shall plot the phase and evolutionary diagrams accordingly. For this reason we must find the critical points of the system of equations (16) forming an autonomous dynamical system and then we linearize the system near the critical points. The system of equations (16)  has the following six critical points for the positive coupling parameter $ \alpha $  i.e. decaying of DE into matter
 :\\

 $\bullet $ I. Critical points: $ P_{1} , P_{2} = ( \pm 1,0,0 )$,\\

 $\bullet $ II. Critical Points : $ P_{3} ,P_{4}= ( -\frac{\lambda}{\sqrt{6}},\pm\sqrt{1-\frac{\lambda^{2}}{6}},0 ) $ \\

 $\bullet $ III. Critical points : $ P_{5} , P_{6} = ( \frac{\alpha-3}{\sqrt{6}\lambda},\pm\sqrt{\frac{\alpha}{3}+ \frac{( \alpha-3 )^{2}}{6\lambda^{2}}},\frac{3-\alpha}{3}( 1-\frac{3-\alpha}{\lambda^{2}} )). $ \\

 where $ \lambda=\frac{V'}{V}=constant $ ,     $  V'=\frac{dV}{d\phi} $.

 On the other hand, for negative coupling parameter we have the following two critical points  \\

 $\bullet $  IV. Critical points : $ P_{7} , P_{8} = ( \pm\sqrt{-\frac{\alpha}{3}},0,1+\frac{\alpha}{3} ) $  where $ \alpha\in ( -3,0 ) $ .\\

 These critical points and the relevant physical parameters at those points are shown in table I and table II.

 \begin{table}
 \caption{ Table shows the location of the critical points and the values of the relevant physical parameters at those points for positive $ \alpha $ .}
 \begin{tabular}{|c|c|c|c|c|c|c|c|c|c|}
 \hline
   $ P_{i} $ & $ x  $ & $  y  $ & $ \Omega_{m} $ & $ \omega_{\phi} $ & $ \omega_{Tot} $ & $ \Omega_{\phi} $ &  $ q $ & $ \omega_{g} $ \\\hline
   $ P_{1} $ &   1    &    0  &         0       &           1        &           1     &            1      &       2 &     0   \\\hline
   $ P_{2} $ &  -1    &    0  &         0       &           1         &           1     &            1      &      2 &  0  \\\hline
   $ P_{3} $ & $ -\frac{\lambda}{\sqrt{6}} $ & $\sqrt{1-\frac{\lambda^{2}}{6}} $ & 0 & $ \frac{\lambda^{2}}{3}-1 $ & $ \frac{\lambda^{2}}{3}-1 $ & 1 & $ -1+\frac{\lambda^{2}}{2} $ & $-1+\frac{\lambda^{2}}{6} $  \\\hline
   $ P_{4} $ & $ -\frac{\lambda}{\sqrt{6}} $ & $-\sqrt{1-\frac{\lambda^{2}}{6}} $ & 0 & $ \frac{\lambda^{2}}{3}-1 $ & $ \frac{\lambda^{2}}{3}-1 $& 1 & $-1+\frac{\lambda^{2}}{2} $ &  $-1+\frac{\lambda^{2}}{6} $  \\\hline
   $ P_{5} $ & $ \frac{\alpha-3}{\sqrt{6}\lambda} $ & $\sqrt{\frac{\alpha}{3}+\frac{(\alpha-3)^{2}}{6\lambda^{2}}}$ &  $\frac{3-\alpha}{3}(1-\frac{3-\alpha}{\lambda^{2}})$ & $ -\frac{\alpha\lambda^{2}}{(\alpha-3)^{2}+\alpha\lambda^{2}} $ &$ -\frac{\alpha}{3}$  &           $ \frac{\alpha}{3} +\frac{(\alpha-3)^{2}}{3\lambda^{2}} $ & $ \frac{1-\alpha}{2}$ & $ -\frac{(3+\alpha)}{6} $ \\\hline
   $ P_{6} $ & $  \frac{\alpha-3}{\sqrt{6}\lambda}  $ & $-\sqrt{\frac{\alpha}{3}+\frac{(\alpha-3)^{2}}{6\lambda^{2}}}$ & $ \frac{3-\alpha}{3}(1-\frac{3-\alpha}{\lambda^{2}})$ & $ -\frac{\alpha\lambda^{2}}{(\alpha-3)^{2}+\alpha\lambda^{2}} $ &$-\frac{\alpha}{3} $ & $ \frac{\alpha}{3}+\frac{(\alpha-3)^{2}}{3\lambda^{2}} $ & $ \frac{1-\alpha}{2} $ & $ -\frac{(3+\alpha)}{6} $  \\\hline
  \hline
 \end{tabular}
 \end{table}

 \begin{table}
 \caption{ Table shows the location of the critical points and the values of the relevant physical parameters at those points when $ -3<\alpha<0 $ .}
 \begin{tabular}{|c|c|c|c|c|c|c|c|c|c|}
 \hline
   $ P_{i} $ & $ x  $ & $  y  $ & $ \Omega_{m} $ & $ \omega_{\phi} $ & $ \omega_{Tot} $ & $ \Omega_{\phi} $ &  $   q   $ & $ \omega_{g} $ \\\hline
   $ P_{7} $ & $ \sqrt{-\frac{\alpha}{3}} $  &  0 & $ 1+ \frac{\alpha}{3}$   &     1        & $ -\frac{\alpha}{3} $    &  $ -\frac{\alpha}{3} $ & $ \frac{1}{2}(1-3\alpha)$ & $ -\frac{(1+\alpha)}{2} $  \\\hline
   $ P_{8} $ & $ -\sqrt{-\frac{\alpha}{3}} $ &    0  & $ 1+ \frac{\alpha}{3}$   &   1         &  $ -\frac{\alpha}{3} $         &   $ -\frac{\alpha}{3} $           &  $   \frac{1}{2}(1-3\alpha)$ & $ -\frac{(1+\alpha)}{2} $     \\\hline
  \hline
 \end{tabular}
 \end{table}
 From table I, we see that the critical points $ P_{1} $ and $ P_{2} $ always exist  ( for all $ \alpha ,  \lambda $ )  while $ P_{3} $ and $ P_{4} $ exist only for $ \lambda^{2} <6 $. Also, the four critical points $ P_{1} $, $ P_{2} $, $ P_{3} $ and $ P_{4} $ represent only the DE components ( DM is absent ). The critical points $ P_{5} $ and $ P_{6} $ exist for $ 2\alpha\lambda^{2} + ( \alpha-3 )^{2}> 0 $ and correspond a combination of DM and DE with the ratio of two energy densities
 $ r=\frac{\Omega_{m}}{\Omega_{\phi}}=\frac{(3-\alpha)(\lambda^{2}-3+\alpha)}{\alpha\lambda^{2}+(\alpha-3)^{2}}  $ .
 For all critical points DE behave like a perfect fluid.
 Further, in table II we see that both the critical points $ P_{7} $ and $ P_{8} $ are combination of DM and DE with the ratio of density parameter
 $ \frac{\Omega_{m}}{\Omega_{\phi}}=-1-\frac{3}{\alpha} $. The equivalent geometric matter behaves as dust for the critical points $ P_{1} $ and $ P_{2} $, it behaves as quintessence matter for the critical points $ P_{3} $ and $ P_{4} $ while the geometric matter acts as phantom fluid for the critical points $ P_{5} $ and $ P_{6} $. \\


 \begin{figure}
 \centering
 \begin{minipage}{.75\columnwidth}
 \centering
 \includegraphics[width=1.0\columnwidth]{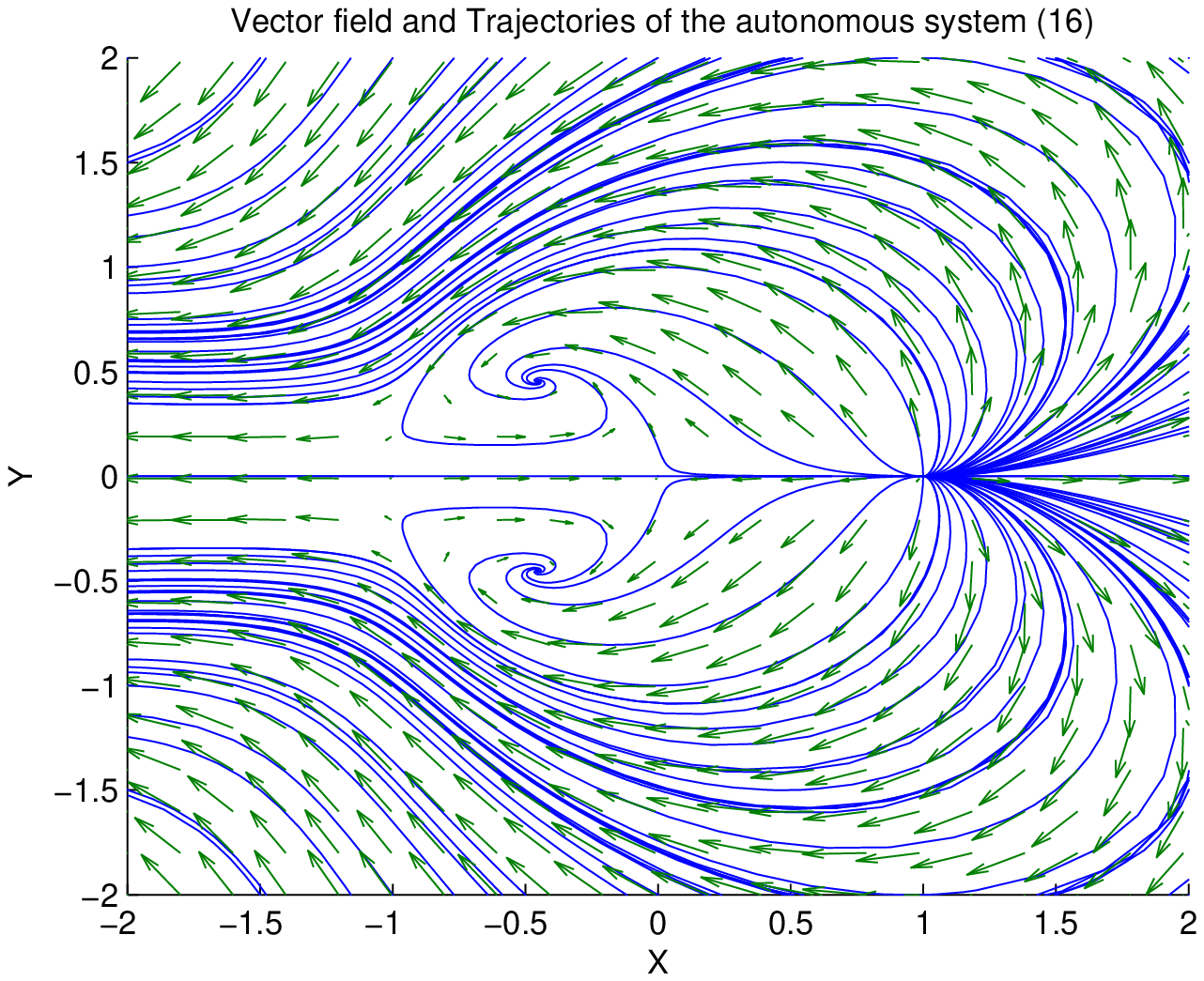}
 \caption{Projections of the phase trajectories  onto the plane ( x, y, $ \Omega_{m} = 0 $ ) of interacting DE in $ f(T) $ model for the choices of
 ($\alpha=0.001 $ , $ \lambda=2.7 $)
  }
  \label{fig:1}
 \end{minipage}%
 \hspace{0.7cm}
 \begin{minipage}{.75\textwidth}
 \centering
      \includegraphics[width= 1.0\linewidth]{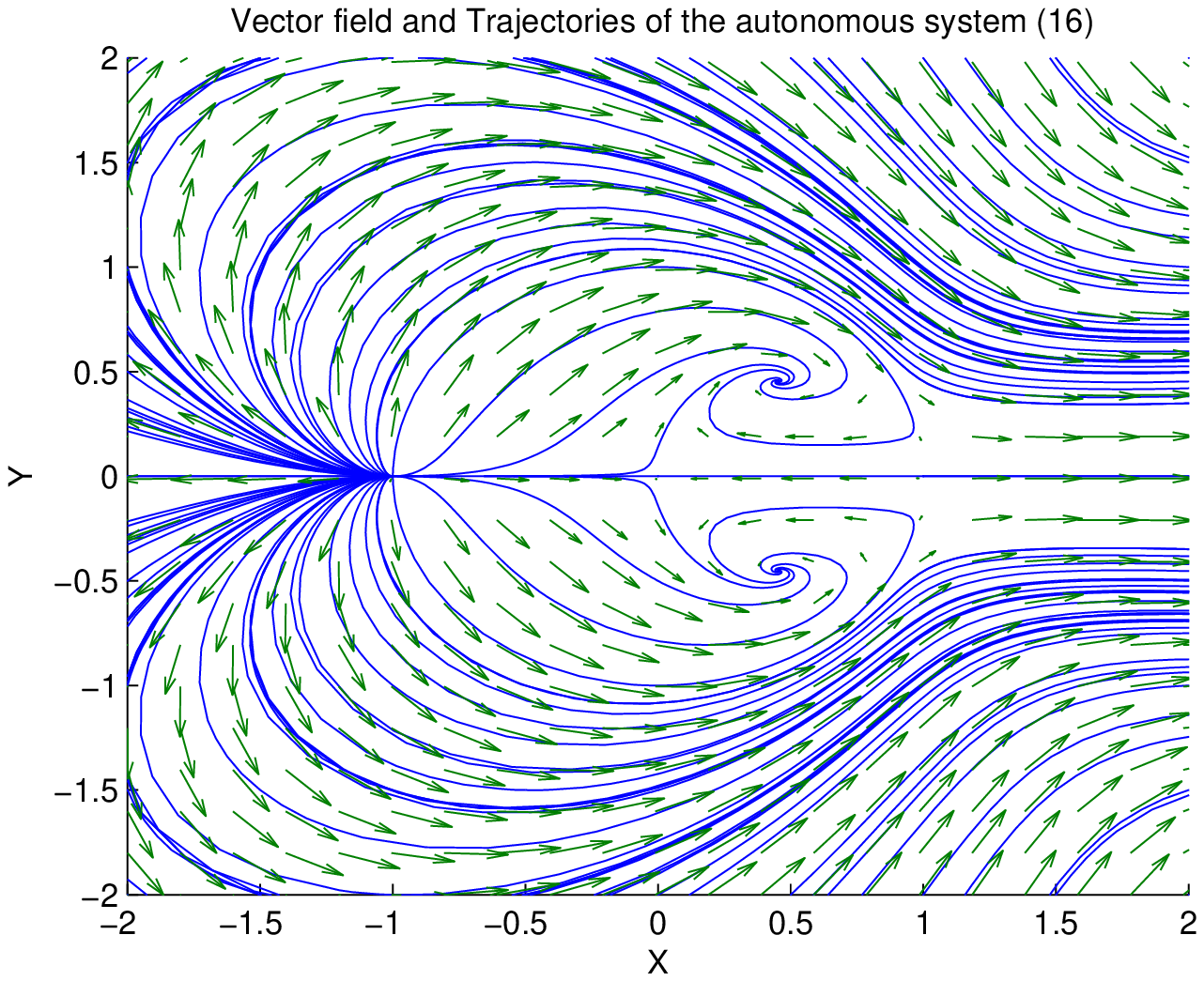}
      \centering
\caption{Phase portrait  of the system (16) for the choices of $ \alpha=0.001 $, $ \lambda=-2.7$.}
 \end{minipage}
 \end{figure}
 \begin{figure}
 \centering
 \begin{minipage}{.5\textwidth}
  \centering
  \includegraphics[width=1.0\linewidth]{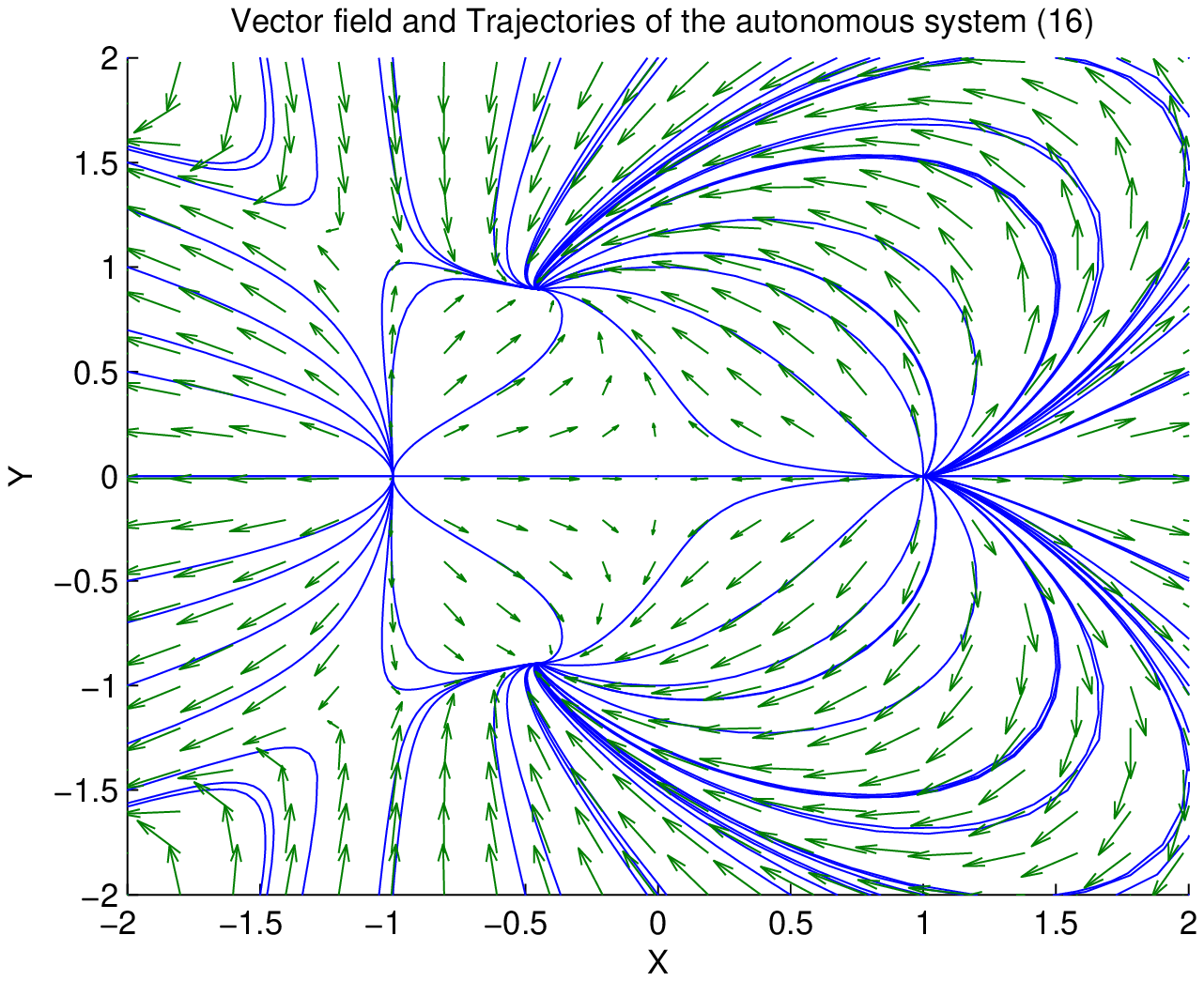}
  \caption{Figure shows the phase portrait of the \\ system (16) for the choices of $\alpha=0.001 $, $\lambda=1.1 $ \\ in the phase plane (x,y,$\Omega_{m}=0 $).}
  \label{fig:test1}
 \end{minipage}%
 \begin{minipage}{.5\textwidth}
 \centering
  \includegraphics[width= 1.0\linewidth]{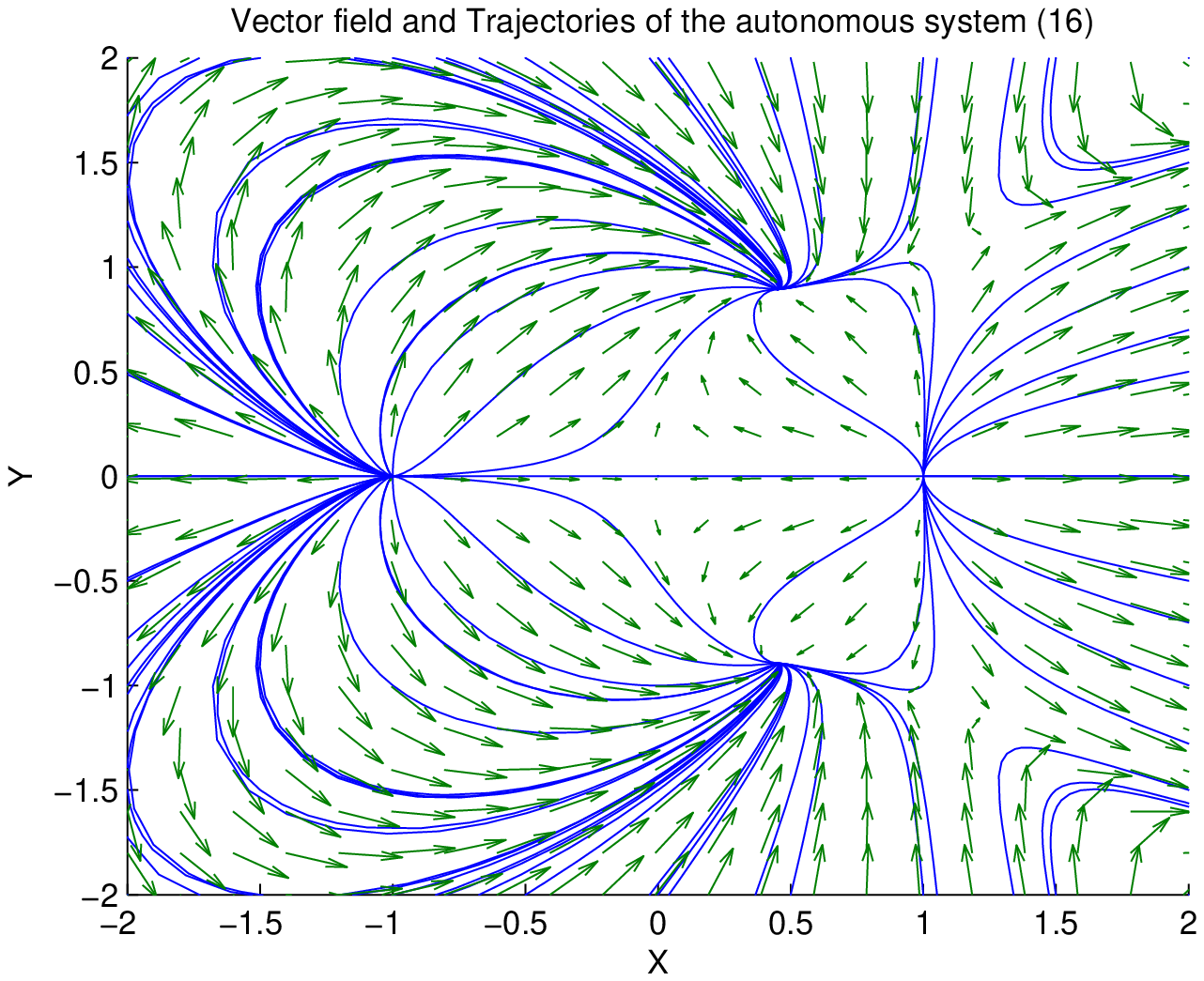}
 \centering
  \caption{Phase portrait of the system (16) for the choices of $\alpha=0.001 $ , $ \lambda=-1.1 $}

 \end{minipage}
 \end{figure}
 \begin{figure}
 \centering
 \begin{minipage}{0.5\textwidth}
  \centering
  \includegraphics[width=1.0\linewidth]{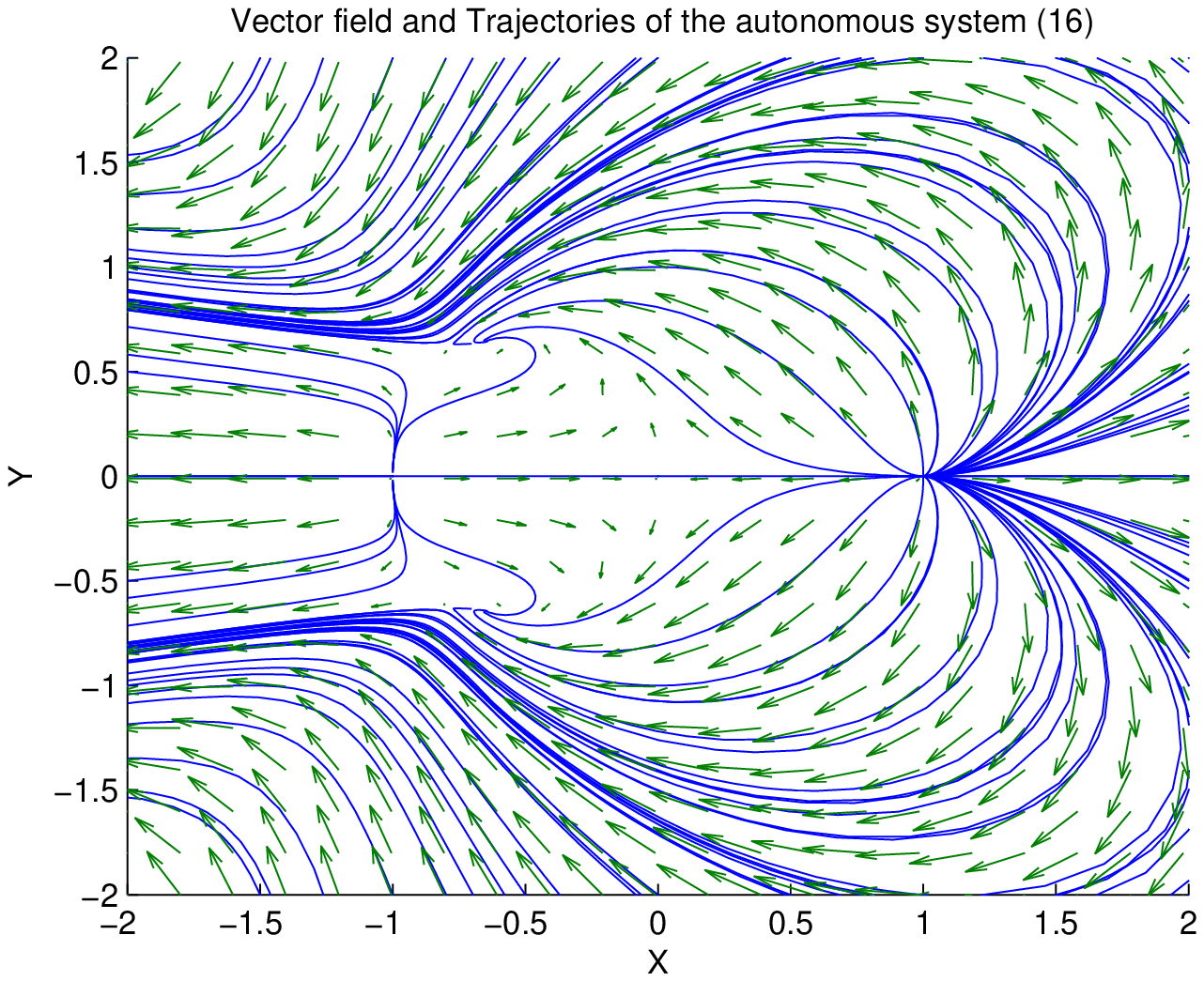}
  \caption{Phase portrait of system of equations (16) \\for the choices of $\alpha=0.001$, $\lambda=1.9$  in the phase \\ plane ( x, y, $ \Omega_{m} =0 $ ) .}
 \end{minipage}%
 \begin{minipage}{0.5\textwidth}
 \centering
  \includegraphics[width= 1.0\linewidth]{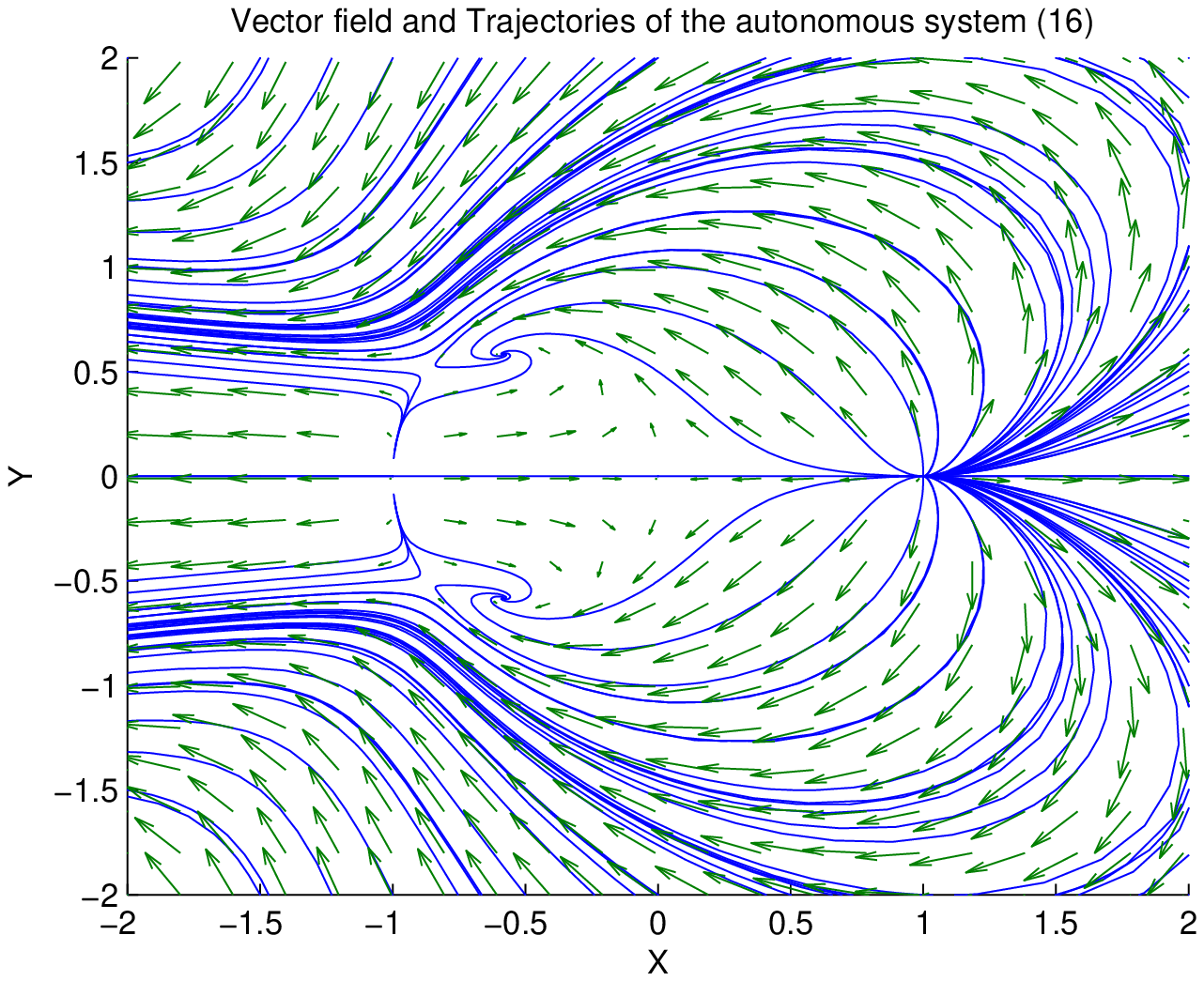}
 \centering
  \caption{Phase portrait for $ \alpha=0.001 $, and $\lambda=2.1$}
  \label{fig:test2}
 \end{minipage}
 \end{figure}
 \begin{figure}
 \centering
 \begin{minipage}{0.75\columnwidth}
 \centering
 \includegraphics[width=1.0\columnwidth]{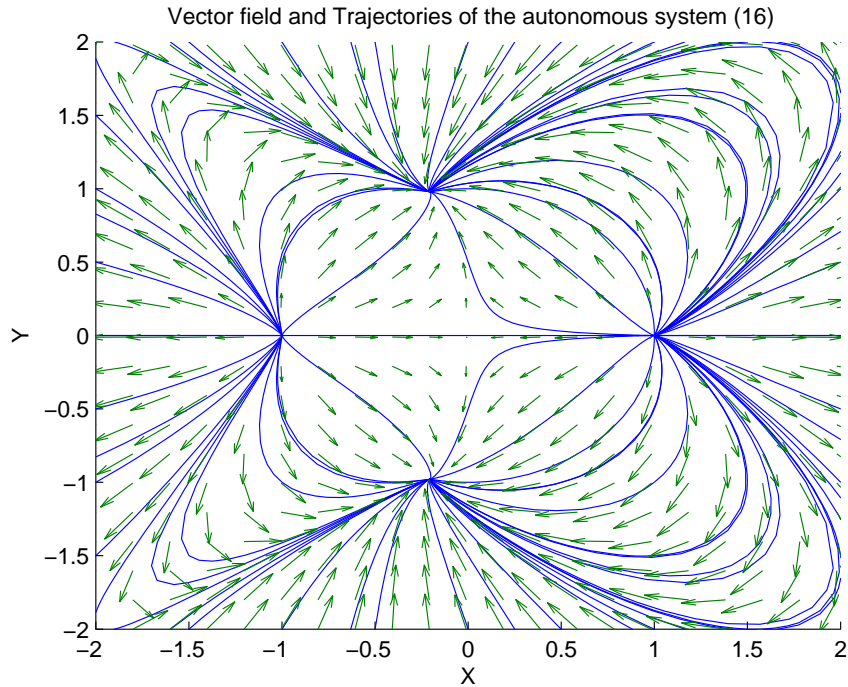}
 \caption{Projection of phase trajectories onto a given plane ( x,y,$ \Omega_{m}=0 $ ) here, we choose $ \Omega_{m}=0 $ . Note that while the point  $ P_{1}$ : (x,y)=(1,0) and the point $ P_{2}$ : (x,y) = (-1,0) are unstable node i.e. source for
 ($\alpha=0.001 $, $ \lambda=0.5 $), the points $ P_{3} $ and $ P_{4} $ (see in table I) are the stable solutions i.e. attractor solutions in the given phase space. Observe that all the trajectories in the phase space always emerge from the points $ P_{1} $ and $ P_{2} $ while all the trajectories enter into the points $ P_{3} $ and $ P_{4} $}
 \end{minipage}%
 \end{figure}

\begin{figure}
 \centering
 \begin{minipage}{0.5\textwidth}
  \centering
  \includegraphics[width=1.0\linewidth]{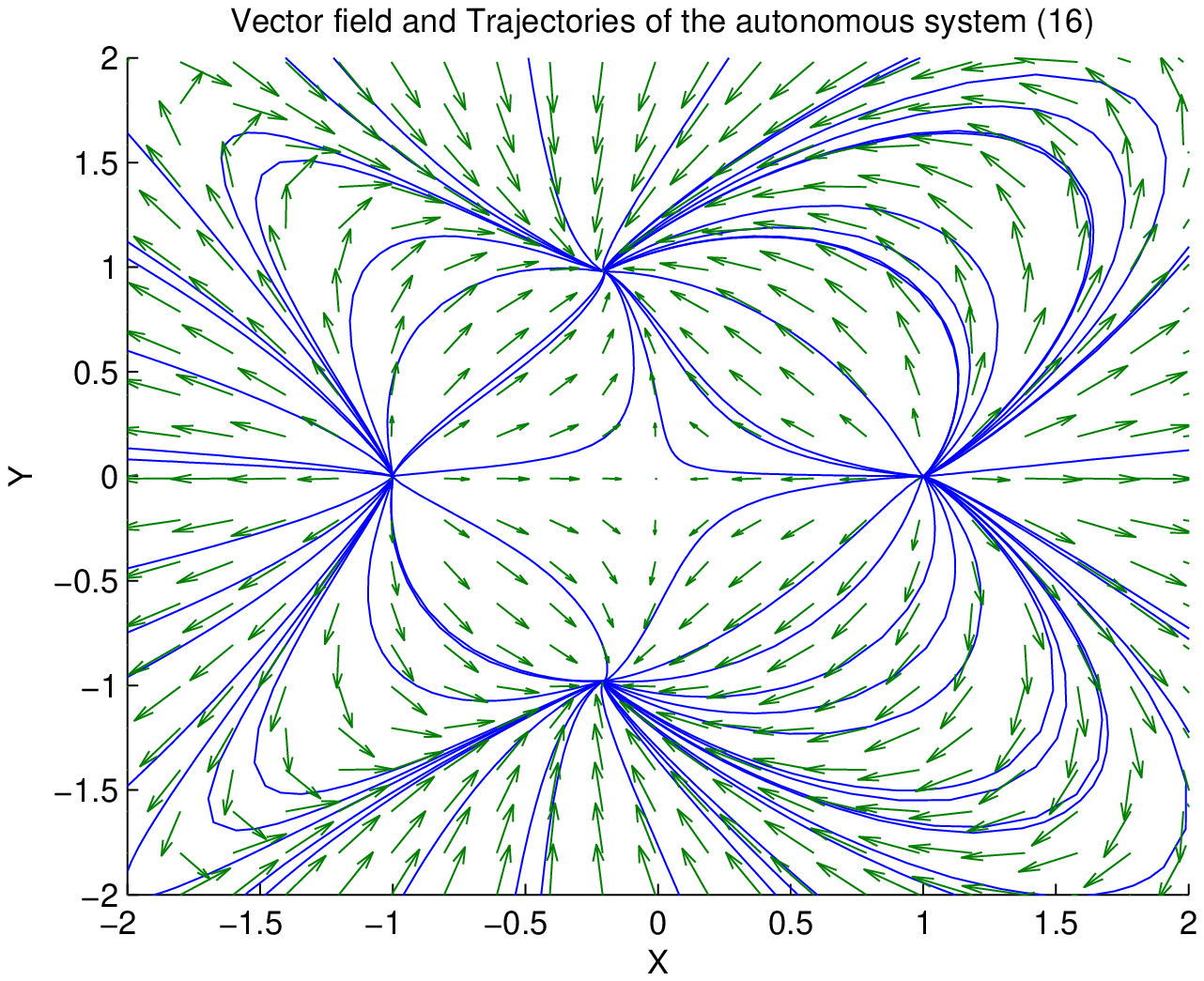}
  \caption{Phase portrait of system of equations (16) \\for the choices of $\alpha=0.001$, $\lambda=0.5 $ with grid \\ difference 0.52}
 \end{minipage}%
 \begin{minipage}{0.5\textwidth}
 \centering
  \includegraphics[width= 1.0\linewidth]{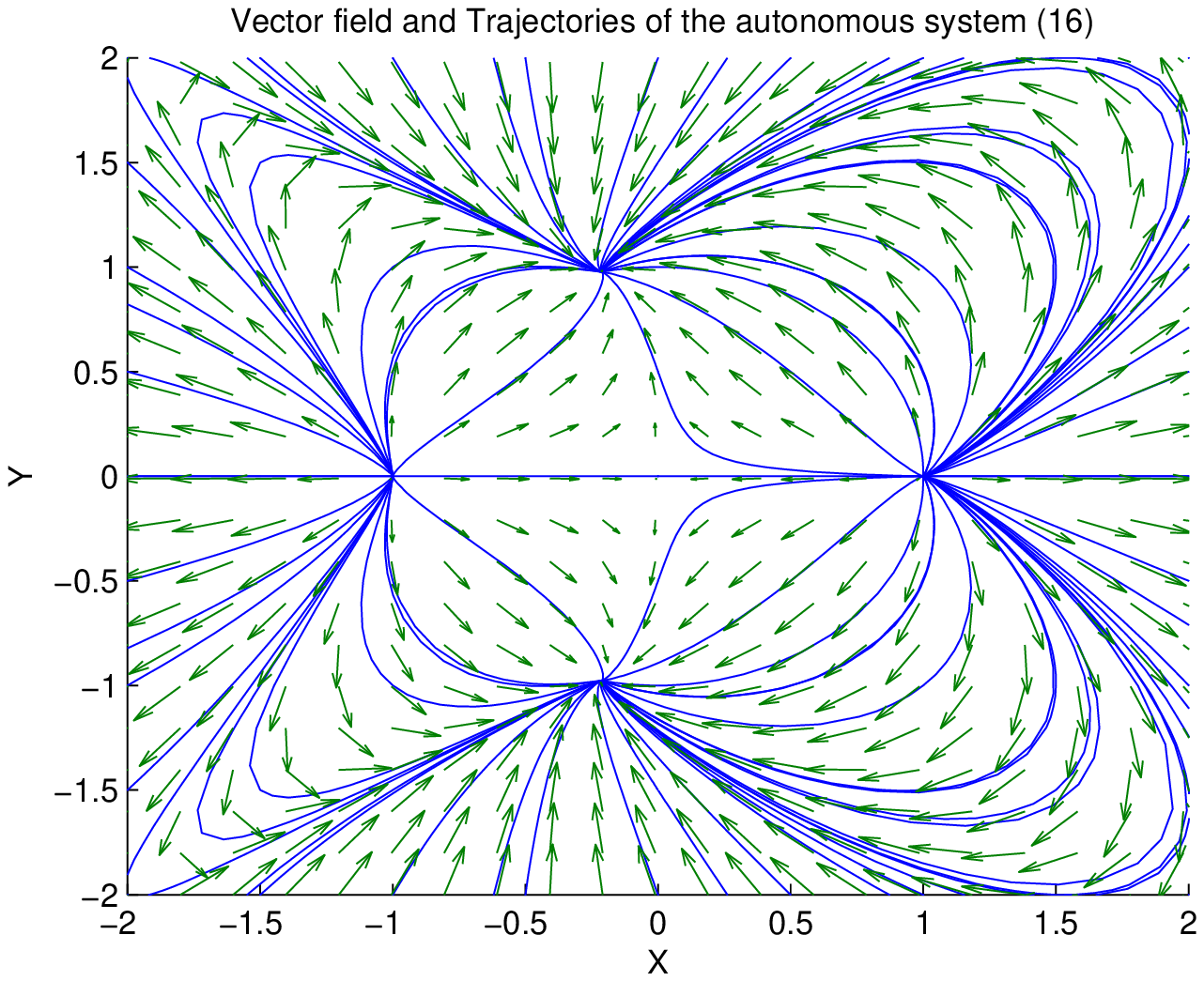}
 \centering
  \caption{Phase portrait for $ \alpha=0.0012 $, and $\lambda=0.52 $ with grid difference 0.5}
  \label{fig:test2}
 \end{minipage}
 \end{figure}

 \section{ Stability Analysis}

 We shall now discuss the stability of the critical points ( presented in Table I and the table II ) of the autonomous system (16), considering first order perturbations near the critical points. To examine the nature of critical points, one has to study the eigenvalues of the first order perturbation matrix which have been presented in Table III and table IV below :  \\
 \begin{table}
 \caption{ Eigenvalues of the linearized matrix for the critical points of the autonomous system (16) when $ \alpha $ is positive and $ A= [(-\frac{3}{2}+\frac{\alpha}{2}+ \frac{7}{4}\frac{(\alpha-3)^{2}}{\lambda^{2}}- \frac{\alpha\lambda^{2}}{\alpha-3})^{2} + \frac{1}{\lambda^{2}}(4\alpha^{2}\lambda^{4} - (\alpha-3)^{4})(\lambda^{2}+\alpha-3)(\frac{1}{\lambda^{2}(\alpha-3)}-3)]^{\frac{1}{2}} $}

 \begin{tabular}{|c|c|c|c|}
   \hline
   $ P_{i} $ & $ \lambda_{1} $ & $ \lambda_{2}  $ & $ \lambda_{3} $ \\\hline
   $ P_{1} $ & $ 3+\alpha $ & $ 3+\sqrt{\frac{3}{2}}\lambda $ & $ 3+\alpha $  \\\hline
   $ P_{2} $ & $ 3+\alpha  $ & $ 3-\sqrt{\frac{3}{2}}\lambda $ & $ 3+\alpha $ \\\hline
   $ P_{3} $ & $ \alpha - 3 + \lambda^{2} $ & $ \frac{\lambda^{2}}{2}-3 $ & $ \alpha - 3 + \lambda^{2} $  \\\hline
   $ P_{4} $ & $ \alpha - 3 + \lambda^{2}  $ & $ \frac{\lambda^{2}}{2}-3 $ & $ \alpha - 3 + \lambda^{2} $  \\\hline
   $ P_{5} $  & 0  & $ \frac{1}{2}[A-\frac{3}{2}-\frac{3\alpha}{2}-\frac{\alpha\lambda^{2}}{\alpha-3}+ \frac{3(\alpha-3)^{2}}{4\lambda^{2}}] $ & $ \frac{1}{2}[-A-\frac{3}{2}-\frac{3\alpha}{2}-\frac{\alpha\lambda^{2}}{\alpha-3}+ \frac{3(\alpha-3)^{2}}{4\lambda^{2}}] $ \\\hline
   $ P_{6} $  &  0  & $ \frac{1}{2}[A-\frac{3}{2}-\frac{3\alpha}{2}-\frac{\alpha\lambda^{2}}{\alpha-3}+ \frac{3(\alpha-3)^{2}}{4\lambda^{2}}] $  & $ \frac{1}{2}[-A-\frac{3}{2}-\frac{3\alpha}{2}-\frac{\alpha\lambda^{2}}{\alpha-3}+ \frac{3(\alpha-3)^{2}}{4\lambda^{2}}] $\\\hline
   \hline
 \end{tabular}
 \end{table}\\
 \begin{table}
 \caption{ Table shows the eigenvalues  of the linearized matrix for the critical points of (16) when $ \alpha \in(-3, 0) $ .}

 \begin{tabular}{|c|c|c|c|}
   \hline
   $ P_{i} $ & $ \lambda_{1} $ & $ \lambda_{2}  $ & $ \lambda_{3} $ \\\hline
   $ P_{7} $ & $ -\alpha-3 $ & $ -\frac{\alpha}{2}+\frac{3}{2}+ \lambda \sqrt{-\frac{\alpha}{2}} $ & 0   \\\hline
   $ P_{8} $ & $ -\alpha-3  $ & $ -\frac{\alpha}{2}+\frac{3}{2}- \lambda \sqrt{-\frac{\alpha}{2}} $ & 0  \\\hline
     \hline
 \end{tabular}
 \end{table}\\
 \begin{table}
 \caption{ Stability criteria of the model where $ B= 6x^{2}+2\sqrt{6}xy^{2}\lambda+\alpha\Omega_{m} \geq 0  $.}
 \begin{tabular}{|c|c|c|c|c|c|c|c|c|c|}
 \hline
   $ C_{s}^{2} $ & For classical stability ($ C_{s}^{2}\geq 0 $) & For quantum stability \\
   &&($ p_{x}\geq 0$ , $ p_{x}+ 2xp_{xx}\geq 0 $ )  \\\hline
   $ 1 + \frac{\sqrt{6}xy^{2}\lambda}{3x^{2}+\frac{\alpha}{2}\Omega_{m}}  $ & $ 6x^{2}+2\sqrt{6}xy^{2}\lambda+\alpha\Omega_{m} \geq 0 $ & $ 6x^{2}+2\sqrt{6}xy^{2}\lambda+\alpha\Omega_{m} \geq 0 $ and \\
   && $ 1 + \frac{4xH}{B}[(6x+ \sqrt{6}\lambda y^{2})\frac{dx}{dN} +  2\sqrt{6}\lambda xy\frac{dy}{dN}  +  \frac{\alpha}{2} \frac{d\Omega_{m}}{dN} ]$ \\
  && $ \leq   9xH (2x^{2}+ \Omega_{m})$ \\\hline
  \hline
 \end{tabular}
 \end{table}

 From table III, for $ \lambda  =-\sqrt{6} $, the critical point $ P_{1} $ is non-hyperbolic in nature otherwise hyperbolic. On the other hand, for $ \lambda =\sqrt{6} $, the point $ P_{2} $ is non-hyperbolic in nature otherwise hyperbolic.\\
 The critical point $ P_{1} $ is saddle node for $ \lambda< -\sqrt{6} $  ( see figure 2. )
 and $ P_{2} $ is saddle node for $ \lambda> \sqrt{6} $  ( see figure 1. )
 while $ P_{1} $ is unstable node for $ \lambda> -\sqrt{6} $  ( fig: 1. )
 and $ P_{2} $ is unstable node for $ \lambda< \sqrt{6} $  ( fig: 2. )
 There are no acceleration phase of the universe near the critical points $ P_{1} $ and $ P_{2} $. Also, these two critical points represent solutions without matter part of the universe. They are only the kinetic energy dominated solutions.

 On the other hand, the critical points $ P_{3} $ and  $ P_{4} $ are same in all respect and are hyperbolic in nature. These two critical points are stable solutions if $ 0<\alpha< 3-\lambda^{2} $ . Also, they behave like attractor solutions for $ \lambda^{2} <3 $  ( i.e. $ -\sqrt{3}< \lambda<\sqrt{3} $ ). But when $ 3< \lambda^{2}<6 $ the two critical points $ P_{3} $ and  $ P_{4} $ behave like saddle node ( these are shown in the phase portrait ). There exists an accelerating phase of the universe near these critical points $ P_{3} $ and  $ P_{4} $ for $ \lambda^{2}<2 $ .

 $ P_{5} $  and $ P_{6} $  are also same in all respect. They are combination of both DM and DE. These solutions are dominated by DE components only for $ \alpha +\lambda^{2} =3 $. $ P_{5} $  and $ P_{6} $ correspond to solutions in accelerating or decelerating phase of universe according as
  $ \alpha  >_{<} 1  $.

  Now, in the table IV, for negative coupling parameter, we get two critical points namely   $ P_{7} $  and $ P_{8} $ which are non-hyperbolic in nature. The solutions are saddle like ( because one eigenvalue is always positive in both of the cases ) due to instability in the eigen direction associated with positive eigenvalue and the stability of an eigen direction associated to a negative eigenvalue. These solutions are dominated by both DM and DE components. There is no accelerating phase near these critical points.

 Thus, our main results can be summarized as follows ( see tables I, III, II and IV ): \\

 $\bullet $   For all  $ \lambda $  and $ \alpha> 0 $, the kinetic energy dominated solutions ( points $ P_{1} $  and $ P_{2} $ in table I ) are always unstable points in phase space. In particular, $ P_{1} $  and $ P_{2} $ represent source or repealer ( unstable nodes ) in phase space for $ \lambda \in(-\sqrt{6} , \sqrt{6} $) and are always decelerating ($ q=2 $) (see figures 3, 4 and 7 ).\\

 $\bullet $   For $ 0<\alpha<3-\lambda^{2} $, $ \lambda^{2}<3 $ ( i.e. $ -\sqrt{3}<\lambda<\sqrt{3} $ ); the scalar field dominated solutions ( points $ P_{3} $  and $ P_{4} $ in table I ) are the late time attractor ( see figures 3 , 4 and 7 ), while for $ 3 <\lambda^{2}<6 $ the solutions are saddle node ( see figures 5 and 6 ). Also, for $ \lambda^{2}< 2 $, there exists an accelerating phase of the universe near the critical points $ P_{3} $ and  $ P_{4} $.\\

 $\bullet $  For all  $ \lambda $  and $ \alpha> 0 $ two critical points $ P_{5} $ and $ P_{6} $ ( see table I ) are non-hyperbolic and are combination of DE and DM components. These are the solutions accelerating or decelerating completely depend on the coupling parameter $ \alpha $. By the linear stability theory, it is not easy to get the stability criteria of these critical points.\\

 $\bullet $  For $ \alpha \in( -3, 0 ) $, the two non-hyperbolic critical points $ P_{7} $ and and $ P_{8} $ ( see in table II ) can be  obtained which are saddle points and are both combination of DE and DM. They are always decelerating ( $ q= \frac{1}{2}(1-3\alpha) $, see table II ). \\

$\bullet $ In the above figures 1-7 the variables $x$ and $y$ are chosen in the range [-2,2] with grid points at a difference of $0.5$. The model parameters $\alpha$ and $\lambda$ are chosen as $\alpha=0.001$ and $\lambda= 0.5$. To examine, whether the present model depends continuously on data we have drawn figures 8 and 9. In fig. 8 we have changed the grid difference to 0.52 while in fig. 9 we have changed the model parameters as  $\alpha=0.0012$ and $\lambda= 0.52$. It is found that both the figures 8 and 9 do not differ significantly from figure 7. So we may conclude that the model depends continuously on data.

 \begin{table}
 \caption{ Condition for stability at each equilibrium point.}
 \begin{tabular}{|c|c|c|c|c|c|c|c|c|c|}
 \hline
   $ P_{i} $ & $ x $ & $ y $ & $ \Omega_{m} $ &  local stability &  classical stability & quantum stability \\\hline

   $ P_{1} $ &   1  &    0 & 0 &  unstable & stable & stable if $ H\geq\frac{1}{18}$ \\\hline
   $ P_{2} $ &  -1  &    0 & 0 &  unstable & stable & stable if$ H\leq -\frac{1}{18} $ \\\hline

   $ P_{3} $ & $ -\frac{\lambda}{\sqrt{6}} $ & $ \sqrt{1-\frac{\lambda^{2}}{6}} $ & 0  & stable (sink) if \\&&&& $ 0 <\alpha<3-\lambda^{2} $ and $ \lambda^{2}<3 $ & stable if $ \lambda^{2} \geq 3 $ & stable if $ \lambda^{2} \geq 3 $

   and
   \\ &&&&&& $ \lambda^{3}H \leq -\sqrt{\frac{2}{3}} $ \\\hline

   $ P_{4} $ & $ -\frac{\lambda}{\sqrt{6}} $ &  $ -\sqrt{1-\frac{\lambda^{2}}{6}} $       & 0 &  stable (sink) if \\&&&& $ 0 <\alpha<3-\lambda^{2} $ and $\lambda^{2}<3 $  & stable if $ \lambda^{2} \geq 3 $ & stable if $ \lambda^{2} \geq 3 $
   and
   \\ &&&&&& $ \lambda^{3}H \leq -\sqrt{\frac{2}{3}} $ \\\hline

   $ P_{5} $ & $ \frac{\alpha-3}{\sqrt{6}\lambda} $ & $ \sqrt{\frac{\alpha}{3}+\frac{(\alpha-3)^{2}}{6\lambda^{2}}}$  & $ \frac{3-\alpha}{3}(1-\frac{3-\alpha}{\lambda^{2}})$  & linear stability fails & stable if $ \alpha \geq 3 $ & stable if $ \alpha \geq 3 $
   and
   \\&&&&&& $(\alpha-3)^{2}\frac{H}{\lambda}\leq -\sqrt{\frac{2}{3}} $   \\\hline
   $ P_{6} $ & $ \frac{\alpha-3}{\sqrt{6}\lambda} $ & $ -\sqrt{\frac{\alpha}{3}+\frac{(\alpha-3)^{2}}{6\lambda^{2}}}$ & $ \frac{3-\alpha}{3}(1-\frac{3-\alpha}{\lambda^{2}}) $  & linear stability fails & stable if $ \alpha \geq 3 $  &   stable if $ \alpha \geq 3 $
   and
   \\&&&&&& $ (\alpha-3)^{2}\frac{H}{\lambda} \leq -\sqrt{\frac{2}{3}} $   \\\hline
   $ P_{7} $ & $ \sqrt{-\frac{\alpha}{3}} $ & 0 & $ 1+ \frac{\alpha}{3} $ & unstable (saddle) & unstable & unstable \\\hline
   $ P_{8} $ & $ - \sqrt{-\frac{\alpha}{3}} $ & 0 & $ 1+ \frac{\alpha}{3} $ & unstable (saddle) & unstable & unstable \\\hline
  \hline
 \end{tabular}
 \end{table}

  \section{ Equilibrium points and Stability criteria }

 In the present three dimensional autonomous system the local stability criteria of an equilibrium point is characterized by the eigenvalues of the perturbation matrix ( presented in table III and table IV for different values of coupling parameter $ \alpha $ ) and then discussion about local stability is presented . We shall now investigate the classical as well as quantum stability of the model.

 In cosmological perturbation, sound speed ( $ C_{s} $ ) has a crucial role in characterizing classical stability. In fact, $ C_{s}^{2} $ appears as a coefficient of the term $ \frac{k^{2}}{a^{2}} $ ( $ k $ is the comoving momentum and '$ a $' is the usual scale factor ) and classical fluctuations may be considered to be stable when  $ C_{s}^{2} $ is positive. On the other hand, for quantum instabilities at UV scale we decompose the scalar field into a homogeneous part ( $ \phi_{0}$ ) and a fluctuation as

                           \begin{equation}
                             \phi(x,t) = \phi_{0}(t) + \delta\phi(x,t)
                         \end{equation}
 Then by expanding the pressure $ p(x,\phi) $, up to second order in $ \delta\phi $, the Hamiltonian for the fluctuations takes the form [48,49]

 \begin{equation}
   \widetilde{H}=(p_{x}+2xp_{xx})\frac{(\delta\dot{\phi})^{2}}{2} + p_{x}\frac{(\nabla\delta\phi)^{2}}{2} - p_{\phi\phi}\frac{(\delta\phi)^{2}}{2}
 \end{equation}

 where suffix stands for differentiation with respect to the corresponding variable[50].

 For positive definiteness of the Hamiltonian we must have

 \begin{equation}
  p_{x}+2xp_{xx}\geq 0    ,    p_{x}\geq 0  ,     - p_{\phi\phi}\geq 0
 \end{equation}\\

 where the first two inequalities are related to quantum stability.\\
 Further, $ C_{s}^{2}$ should be less than unity. Because otherwise it is possible to send signals along space-like world lines and this would open a Pandora's box of classical time travel Paradoxes and this also violates the unitarity of quantum theory. Further, if  $ C_{s}^{2}>1$ then from the effective field theory the fluctuations $ \delta\phi $ propagate faster than the speed of light and therefore there will be violation of Causality.
 In the present cosmological scenario we have

\begin{equation}
        C_{s}^{2} = 1 + \frac{\sqrt{6}xy^{2}\lambda}{3x^{2}+\frac{\alpha}{2}\Omega_{m}}
\end{equation}

 so for classical stability ( assuming $ \alpha > 0 $ )

\begin{equation}
            6x^{2}+2\sqrt{6}xy^{2}\lambda +\alpha \Omega_{m} \geq 0
\end{equation}

  while for quantum stability the restrictions are

        $$ 6x^{2}+2\sqrt{6}xy^{2}\lambda +\alpha \Omega_{m} \geq 0 $$
                  $$ and $$
        $$   1 + \frac{4xH}{B}[(6x+\sqrt{6}\lambda y^{2} )\frac{dx}{dN} + 2\sqrt{6}\lambda xy \frac{dy}{dN} + \frac{\alpha}{2} \frac{d\Omega_{m}}{dN} ] \leq 9xH(2x^{2} + \Omega_{m})$$ \\

  where $ B = 6x^{2}+2\sqrt{6}xy^{2}\lambda +\alpha \Omega_{m}  $   which is positive, H is the Hubble parameter and the values of $ \frac{dx}{dN} $ , $ \frac{dy}{dN} $ and         $ \frac{d\Omega_{m}}{dN} $   are given in the system of  equations (16).\\

  We have shown both classical and quantum stability criteria of the model in table V. We shall now discuss about the criteria for the model stability at the equilibrium points ( presented in table I and table II ) when x,y and $ \Omega_{m} $  take the corresponding values at the equilibrium points. From the tables I and III we see that the equilibrium points $ P_{1} $ and $ P_{2} $ are not locally stable( saddle node or unstable node ), they behave like source for $ \lambda \in ( -\sqrt{6},\sqrt{6} ) $ and from the above model stability analysis they correspond to classical stability only but conditional  quantum stable i.e $ P_{1} $ corresponds to quantum stability if $ H \geq \frac{1}{18} $ while for $ H \leq -\frac{1}{18} $, the critical point $ P_{2} $ corresponds to quantum stability. As the equilibrium points $ P_{3} $ and $ P_{4} $ are same in all respect so they correspond to classical stability of the model if $ \lambda^{2} \geq 3 $ while for quantum stability  the restrictions are   $ \lambda^{2} \geq 3 $ and $ \lambda^{3}H \leq -\sqrt{\frac{2}{3}} $ ( see table VI ). From local stability analysis, we see that the critical points $ P_{3} $ and $ P_{4} $ ( table I ) are stable if  $ \lambda^{2} < 3 $. From the above model stability analysis we see that the two equilibrium points  $ P_{5} $ and $ P_{6} $ are classical stable if $ \alpha \geq 3 $ where as they correspond to quantum stability if $ \alpha \geq 3 $ and $ ( \alpha-3 )^{2} \frac{H}{\lambda} \leq - \sqrt{\frac{2}{3}} $ ( presented in table VI ). Finally, from the tables II and IV, we see that two equilibrium points $ P_{7} $ and $ P_{8} $ are not locally stable (saddle) and from the above analysis we see that they are not classical as well as not quantum stable. The corresponding condition for stability at each equilibrium point are presented in table VI. \\

  \section{ Equilibrium points and Cosmological implications }

 From the above phase space analysis of the $ f(T) $ model we shall now discuss about the cosmological behavior of the model at the equilibrium points. From table I, we find that $ P_{1} $ and $ P_{2} $ indicate the universe is completely dominated by kinetic energy of the scalar field and late time acceleration is not possible. They correspond to flat non-accelerating unstable universe without matter part and so they are not of much interest in the present context.

 On the other hand, the  points $ P_{3} $ and $ P_{4} $ are interesting from the cosmological point of view (for $ \lambda^{2}<2 $). There exists an accelerated expansion of the universe near $P_{3}$ and $ P_{4} $ which correspond to DE model in the quintessence era and the critical points are stable as well as the model is locally stable. It should be mentioned that these critical points correspond to cosmological solutions which describe the recently observed late time acceleration of the universe ( details are shown in figures ). Thus, our model predicts the correct evolutionary scheme for DE density in the regime of this quintessence scalar field. However, for the restriction $ 2<\lambda^{2} <3 $ the points $ P_{3} $ and $ P_{4} $ are stable from local analysis and at these points the potential energy dominates over the kinetic part and the scalar field behaves as exotic fluid. The equilibrium points $ P_{5} $ and $ P_{6} $ are the potential energy dominated non hyperbolic solutions so we cannot investigate the local stability criteria of the system by the linear stability theory. They are the combination of both DM and DE. Accelerating or decelerating phase of the universe near the points $ P_{5} $ and $ P_{6} $ are fully depended on the coupling parameter of the interaction term of DM and DE. The equilibrium points $ P_{7} $ and $ P_{8} $ are same in all respect for the negative coupling parameter. These are the kinetic energy dominated solutions of the scalar field and there exists an non-accelerating universe near $ P_{7} $ and $ P_{8} $.

 \section{ Discussion and  Concluding Remarks }

 In this paper we have performed a dynamical system analysis of a complicated model based on an alternative theory of gravity- the so called $ f(T) $ gravity. The matter is chosen as dark species - dark matter and dark energy which are represented by dust and a scalar field with an exponential potential respectively. We have investigated the stability and phase space description of interacting dark energy in $ f(T) $ cosmology by introducing a simple interaction between dark energy and the  matter  content  of the  universe. We have written the general dynamical system equations from basic evolution equations by the suitable variables which are normalized over Hubble scale and have found two sets of critical points for positive and negative coupling parameter ( see table I and II ). In the context of $ f(T)$ gravity the negative coupling parameter shows that the energy is being transferred from matter to dark energy. We get a finite phase space of the system of evolution equations ( autonomous system ) and phase space forms a paraboloid bounded by $ \Omega_{m} = 0 $ and $ \Omega_{m} = 1 $. From the phase space analysis and  stability analysis of the critical points  ( presented in tables I and II ) we find the stable and unstable solutions ( see figures 1-7 )  for the different values of parameters involved. Also the accelerating or decelerating behaviour of the universe near both the hyperbolic and non-hyperbolic critical points have been given in the last section. It is worth noting that the recently observed accelerated expansion of the universe near the equilibrium points $ P_{3} $ and $ P_{4} $ can be seen from the autonomous system (16) in the quitessence scalar field. These support the recent evolutionary scheme of the universe.\\

Further, we have investigated the classical as well as quantum stability of the model. Note that these two types of stability are not interrelated because the stability of the critical point  is related to the perturbations $ \delta x $ , $ \delta y $ and $ \delta \Omega_{m} $ ( the corresponding variables of critical point ). On the other hand, the classical stability of the model is connected to the perturbations  $ \delta p $ (and depends on the conditions $ C_{s}^{2}\geq 0 $ ) while the quantum stability is related to the perturbations $ \delta \phi $ and the conditions take the form of inequalities (24). Thus the critical points can be classified into three categories namely  \\\\
 (i) unstable points at which the model is stable .\\
 (ii) stable points at which model is unstable  and \\
 (iii) stable points with stable ( both classical and quantum ) model.\\\\
 From the table I, the equilibrium points $ P_{3} $ and $ P_{4} $ are interesting in the present context, they purely describe the scalar field dominated solutions in the $ f(T) $ model and describe the late time acceleration. So, we conclude that although the equilibrium points $ P_{3} $ , $ P_{4} $ correspond to stable model configuration but locally they are not stable equilibrium points. However, it should be noted that the dimensionless variables introduced in equation (15) are not suitable for a complete analysis. Though, the analysis of the static solutions that correspond to equilibrium points at infinity could be analyzed by Poincare projection technique, but for bouncing scenario the expansion rate may pass through zero making the state space non-compact. Therefore, for future work, it is desirable to choose appropriate compact variables to care of expansions, collapses, static solutions and bounces as well.


\begin{acknowledgments}

    One of the authors (S C) is thankful to UGC-DRS programme , in the Department of the Mathematics, J.U . S C is also thankful to IUCAA , Pune for research facilities at Library.

\end{acknowledgments}


%

\begin{thebibliography}{}
\bibitem{REf}\label{50} {Usually , the fluctuations of a time - varying scalar obey Lorentz - violating dispersion relations. However , in cosmological context , Lorentz invariance is always violated considering preferred ( CMB -, comoving observers-) frame. }
\end{thebibliography}
\end{document}